# On the Design and Study of an Installation for Office Workers to Amplify Temporal Diversity and Connection to Nature

Josh Andres [1], Rodolfo Ocampo [1,2], Hannah R. Feldman [1], Louisa Shen [1], Charlton Hill [3], Caroline Pegram [1,3], Adrian Schmidt [1], Justin Shave [3], Brendan Wright [3]

[1.] School of Cybernetics, The Australian National University, Australia
[2.] University of New South Wales Faculty of Art & Design, Interactive Media Lab, Australia
[3.] Uncanny Valley Technology Music Company, Australia.
[firstname.lastname]@anu.edu.au, [firstname]@uncannyvalley.com.au

## Abstract

We present the design and user study of an installation for office workers, enabling moments of temporal diversity and connection to nature. The installation is a form of creative computing experience that departs from the traditional focus on office technologies for productivity. Drawing on neuroscience insights and the slowing effect of nature sounds on time perception, we created an immersive, slow interaction, generative AI installation that composes an audiovisual space — serving as a perceptual portal into temporal realms beyond the linear rhythm of the office. Our study investigates the lived experiences of 18 office workers, gathered via explicitation interviews, observational notes, and video recordings, analysed through an inductive thematic analysis. Key findings highlight the ephemeral qualities in creative computing experiences using generative AI, its potential to foster contemplative practices, amplify ecological temporalities, and reshape office workers' engagement with their environment. Our design and user study offer research and practical implications for utilising creative computing to enrich office experiences.

## Introduction

Office technologies have traditionally aimed to boost efficiency and productivity, shaping work practices and spaces in a dynamic interplay (Hayles, 2012, 2017). These tools usually emphasise utility, often missing chances for playful exploration and moments of nature contemplation that could support workers' wellbeing.

Recently, integrating nature-inspired elements in offices has positively influenced workers' experiences (Colenberg et al., 2021; Kaplan, 2007; Mayer et al., 2009). These elements, typically peripheral like décor, organic architecture, and natural light, contrast with creating sensory engagements to foster connection to nature and temporal diversity.

We present an experimental immersive installation that frames technology beyond a tool that augments workers' productivity to a medium for fostering moments of slow playful exploration and nature contemplation. The installation departs from the sedentary experience associated with office work and invites slow movement via gestural interaction to engage with an audiovisual experience representing multiple time scales. These time scales comprise various data streams from within and around the office, such as social media, commerce, infrastructure, governance, culture, and nature. They range from rapidly changing to those that change over decades and centuries, offering a glimpse into multiple temporal perspectives. The interactive audiovisual experience encourages office workers to engage with the unpredictability and variability of nature and human activity, contrasting with the structured rhythms of office life. It enhances their ability to perceive multiple dimensions of time, broadening their temporal awareness and connection to their surroundings.

We investigate the lived experience of 18 volunteer office workers through explicitation interviews, observational notes, and video recordings via an inductive thematic analysis. Resulting in four themes with nine subthemes and four design implications on creative computing technologies that enable playful, nature-contemplative moments as wellbeing-enriching opportunities in the office.

## Related Work

**The Emergence of Office Worker Digital Technologies**
In the 1950s, office technologies like dictaphones, photocopiers, and printers began to shape working rhythms and streamline tasks (Ong, 2012; Yates, 1993). The following decade brought further advancements. Gordon Pask's Self-Adaptive Keyboard Instructor (SAKI) enhanced industrial perceptual motor skills, facilitating worker training (Bird & Di Paolo, 2008; Pask, 1963). Engelbart's demonstration of augmenting human intellect showcased how digital technologies could be seamlessly integrated into workflows to boost productivity (Engelbart, 2023). And Licklider's work on human-computer symbiosis described the challenges for creating effective partnerships between workers and computers, aiming to surpass individual cognitive abilities (Licklider, 1960). These works pioneered new approaches to thinking, learning, and collaborating via technology. Our work also aims to support humans through technology by enabling enriching experiences and envisioning the future of creative computing for offices.

**Natured-Inspired Offices for Sensorial Enrichment**
The early 21st century saw a resurgence of biophilic design in office spaces, emphasising the benefits of and human affinity for nature (Arvay, 2018; Joye & Van den Berg, 2011). This approach integrated nature, such as plants, and natural elements, such as lighting and organic textures, to positively influence worker wellbeing and reduce potential sensory deprivation in digital and machinery-dominated environments (Downton et al., 2017; Kaplan, 2007). Interestingly, while the integration of nature into office spaces positively influences workers' wellbeing, it has mainly focused on peripheral aspects, such as lighting and decorative features. Our work seeks to expand on the benefits of integrating nature by facilitating moments of nature contemplation through creative computing approaches, wherein nature elements become the primary focus of attention.

**Worker Augmentation in The Future of Work**
The continued advancement of AI and IoT technologies has sparked discussions about their impact on worker autonomy and privacy (Desjardins & Wakkary, 2016; Rahwan, 2018). While these technologies have streamlined operations, they also present challenges such as the risk of excessive automation and potential sensory dullness in workspaces. The discourse on the future of work has largely centered on enabling remote work through technology, which, while increasing flexibility for some, has also made it difficult for others to disconnect from work, leading to higher stress levels (Furendal & Jebari, 2023; Teevan et al., 2022). Similarly, the discussion about AI in the office seeks to create office workers and AI partnerships for higher productivity (Ashktorab et al., 2021; Wang et al., 2021; Weidele et al., 2020) . As we delve into the future roles of technology in work, our research speculates on the types of creative computing experiences that could support workers beyond just enhancing productivity. This underscores the importance of balancing technological integration with sensory and experiential richness. It aligns with our efforts to create playful and temporally diverse office environments that enhance office workers' experiences through creative computing.

**Worker Sensory Enrichment Installations**
Previous works investigated the evolution of the office with the introduction of technology (Andres, Ocampo, et al., 2023; Botin & Hyams, 2021; Duffy et al., 2018), focusing on opportunities for sensory enrichment. Studies have explored how energy usage can be represented, for instance via eco-visualisations (Holmes, 2009; Pierce et al., 2008) and renewable energy (Pierce & Roedl, 2008; Törnroth et al., 2022). These efforts illustrate the dynamic interplay between office workers, the office building, and their combined ecological footprint, framing this relationship through the lens of energy utilisation.

Sensory enrichment installations have also explored the collective mood of the office as an aggregated representation. The "cloud mood" installation (Scolere et al., 2016) collected and visually represented the emotional state of the workplace, opening opportunities for empathy and conversation. Similarly, the "mood squeezer" (Gallacher et al., 2015) reflects the collective mood through dynamic colour projections to increase awareness of the office's emotional climate. These installations represent the collective relationship of office workers through the temporal perspective of changing moods. Importantly, collective emotional representations can be unwelcome in contemporary offices due to office workers' concerns over employers' sensing their information and the regulation of AI emotion sensing technologies (Ajunwa et al., 2017; Boyd & Andalibi, 2023).

These works highlight the intricate nature of office worker relationships, and the role technology plays in amplifying or masking information. They inspire us to explore how creative computing experiences can be designed to promote positive social interactions in the workplace.

**Research Gap**
Existing research on HCI in the office primarily focuses on supporting office workers' productivity and the aesthetic quality of office spaces. While nature-inspired elements in offices have been positively linked to worker wellbeing, their application often remains limited to peripheral or decorative aspects rather than providing immersive, multi-sensory experiences that could deepen engagement with nature.

There is a research gap in how to design creative computing technologies that augment office workers' temporal experiences through playful and nature-contemplative moments as wellbeing enriching opportunities. Our work contributes to this gap by designing, developing, and studying a dynamic generative installation that emphasises sensory enrichment, temporal diversity, and deeper connection to nature over technologies for productivity in an office context.

## Methodology
We investigate a future form of creative computing experiences for offices using "Research through Design" (RtD) (Zimmerman et al., 2007; Zimmerman & Forlizzi, 2014) and a first-person explicitation approach (Vermersch, 1994). Within RtD, our installation serves as a speculative "physical embodiment" artefact that offers new knowledge about the design implications and potentials of creative computing for offices (Gaver et al., 2004; Matthews, 2014), as discussed in the section Discussion and Implications.

To understand office worker's experiences and the futures the installation evokes, we conducted a user study using the explicitation approach (Vermersch, 1994). This approach facilitates participants' transition from a pre-reflective initial state to a more analytical articulation of their encounters with technology. Alongside observational notes, images, and video recordings, this approach enabled a comprehensive analysis of their experiences through an inductive thematic analysis, described in the User Study and Findings. Finally, the user study provoked world accounts beyond participants' inquiry into the installation itself. Serving as a stepping stone to encourage their reflection and inspiration about future creative computing experiences for offices presented in the section Discussion and Implications.

## Design Process

### Representing Multiple Time Scales

In developing our installation, we were guided by a core concept: creating a creative computing experience that enables office workers to experience multiple time scales as an immersive, playful, nature-contemplative experience.

We draw from neuroscience research showing that the audio and visual information we perceive, along with our motor movements, influence our perception of time (Craig, 2002; Tajadura-Jiménez et al., 2015; Wittmann, 2009). For instance, slow movements can decelerate our perception of time (De Kock et al., 2021; Yon et al., 2017), leading to emotional and visceral moments that mark our temporal experience. Additionally, nature elements, including sounds, have a slowing effect on our physiology, positively influencing human wellbeing (Davydenko & Peetz, 2017; Duffin, 2023; Jo et al., 2019).

To represent multiple time scales, we drew inspiration from Brand's pace layer model (Brand, 2018), conceptualising six-time scales as relations between human activity, technology, and the natural environment. The scales range from changing in seconds to changing over decades and centuries; these are, social media (adapted from fashion), commerce, infrastructure, governance, culture, and nature. We contextualise the pace layer model to our office by assigning an API to each time scale as detailed in Table 1 below.

| Time scale | Representing data |
| --- | --- |
| **Social Media**: This pertains to rapidly changing information, from seconds to minutes, as exemplified by social media content produced by office workers. | The latest tweets from School of Cybernetics Twitter handle. |
| **Commerce**: This involves the daily and monthly economic activities of the city & state of the office. | Latest Economy and Growth indicators for *ACT and AUS* from the WorldBank open data API |
| **Infrastructure**: This relates to the evolution of physical and digital infrastructures, like roads and telecommunications, changing over months and years in the city and state where the office is located. | Latest Infrastructure indicators for *ACT and AUS* from the WorldBank open data API |
| **Governance**: This encompasses the local policies and building management processes, which adapt over several years. | Latest Public Sector indicators for *ACT and AUS* from the WorldBank open data API |
| **Culture**: This reflects the city's and country's interests over decades, illustrated by the most-read Wikipedia articles and popular topics. | Most read Wikipedia articles of the day from *ACT and AUS* |
| **Nature**: This refers to the natural environment surrounding the office, which changes over decades and centuries. | Global $CO_2$ concentration in the atmosphere (ppm) per month for the last 36 months - updated weekly from *Canberra* |

Table 1: The Pace Layers contextualised to the office workers, the office space, and natural environment via various data sources.

### Audio and Visual Representations

We developed a system, shown in Figure 1, that uses GPT-3 (Floridi & Chiriatti, 2020) to process data streams. It prompts GPT-3 to "explain the data," generating a sentence and labels describing its emotional properties for each time scale, as detailed in the next sections.

Our system contains 1,050 professional musical compositions created for this work, including nature sounds from the local surroundings. We invited a diverse group of people from across and outside our institution, including academics, professionals, students, and retirees (N = 80), to individually label the sounds via a custom-built interface based on the emotions they experienced. To improve the coverage of labelled sounds, the interface randomly loaded sounds to each labeller. Only 219 sounds of 1050 received more than one label. The labels averaged 2.9 words per label. The soundscape is generated by semantically matching the labels from each time scale sentence with those created by the office workers. We partnered with artificial intelligence music company Uncanny Valley and built on their generative music engine to create the soundscape, as detailed below.

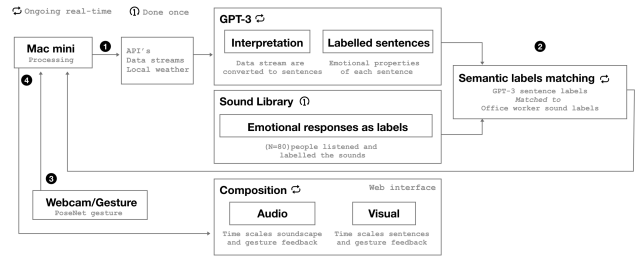

Figure 1: System architecture. 1: Data streams across various time scales are transformed into sentences and labels reflecting emotional properties. The sound library is labeled by office workers based on their emotional responses. 2: These labels are semantically matched to initiate sound generation. 3: Gestural interactions provide real-time audiovisual feedback. 4: Displays the ongoing audiovisual composition.

### From Data Streams to Sentences and Labels via GPT

We followed (Ocampo et al., 2023) GPT-3 semantic data sonification and prompt engineering approach, which includes the evaluation of prompts to generate from data streams sentences and labels representing their emotional qualities. We used the prompt `"data explanation"` on Engine: text-DaVinci-002 to generate the time scale sentences. Subsequently, we used the prompt `"Identify three words capturing the data's emotional essence"` to analyse the resulting sentence and derive labels representing its emotional quality.

Our intention in leveraging the benefits and limitations of GPT-3 (Floridi & Chiriatti, 2020) to `"explain the data"` was to create temporal representations of each data stream that are ambiguous, witty, contradictory, and insightful. These diverse representations highlight the value of experiences that can "stimulate critical reflection" and engagement (Bardzell et al., 2014; Odom et al., 2012).

**Soundscape Generation via Semantic Comparison**
Using GPT-3, we converted each data stream into text sentences, resulting in a 1024-vector embedding. The sentences, including their associated emotion labels, were updated daily, and through semantic comparison via cosine similarity calculations between word embeddings (Rahutomo et al., 2012), using OpenAI's "text-similarity-ada-001" model, were paired with the labels created by the office workers. Resulting in emotional temporal qualities from the sentences matched to the emotional qualities experienced by the office workers when listening to the sounds. These paired labels served as the seed to prompt the generative music engine to create the soundscape. The soundscape is designed to be of allusive quality, referring to data sonification that bridges complex information with musical expression for sensory and emotional experiences (Mardakheh & Wilson, 2022). The generative music engine ensures musical congruence, referring to the soundscape's melody, harmony, rhythm and tempo, by 1) randomly assigning a global music key, 2) assigning a musical mode using the local weather via the weather API, where sunny weather results in a more upbeat mode, rainy weather in a melancholic mode, and partly cloudy is neutral mode), and by, 3) assigning each time scale a musical attribute as follows: Social media uses Accents, Commerce – Bass, Infrastructure – Timekeepers, Governance – Pads and chords, Culture – Scales, Nature – Atmospheric.

**Slow and Contemplative Gestural Interaction**
The installation's interface is designed to promote slow, contemplative engagement. It uses three slow design qualities inspired by Odom et al (Odom et al., 2022).

"Temporal Interconnectedness", to represent multiple time scales that enable the "possibility for temporal interconnectedness to emerge" through the visuals and the soundscape.

"Temporal Drift", where the device's ongoing representation of time departs from that of the user's daily cycle, in our case, varying in months, decades, and centuries through the multiple time scales represented.

"Explicit Slowness", where the user has restricted control over changing the device's pace; in our case, the user can not alter the pace of change in each time scale or the generation of the audiovisuals.

Office workers interact via gestural interaction, their hands are represented on the screen by two white circles; the circles leave a fading trail of nature-inspired colours as they move (greens, browns, yellows, blues). The white strings separating each time scale vary in thickness and vibration motion (see Figure 2), representing each time scale's temporal quality. For example, the first, fast-changing time scale, has a more volatile vibration, while the last, slow-changing, is steady and gradual. The white circles crossing a string produce a musical note (top: A, D, G, B, and E), inviting slow interaction implicit in the prototype.

We implemented gesture interaction with Posenet (*MoveNet & TensorFlow.Js*, n.d.), using heatmaps and processing data locally to ensure participants privacy. The interface uses a 65-inch 3K TV screen connected to a Mac mini to render the graphics and play the soundscape via a speaker bar to offer an immersive sound experience. A standard webcam facilitates gesture interaction, set at a 1-meter height to cater for people standing or in a wheelchair.

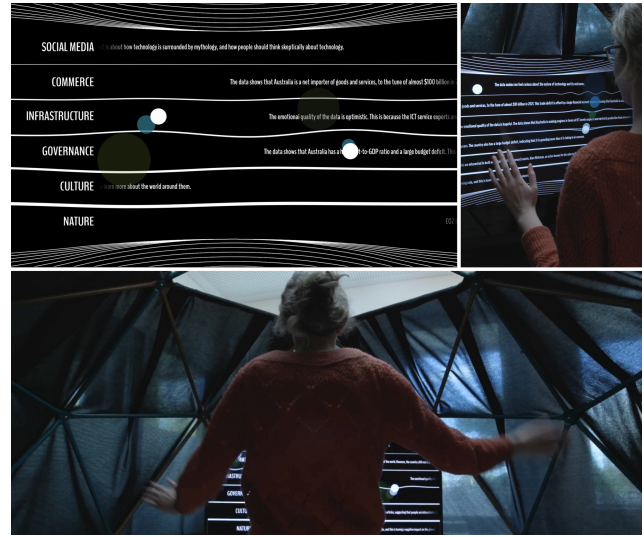

Figure 2: An image of the installation's interface and a participant interacting with the audiovisual time scales.

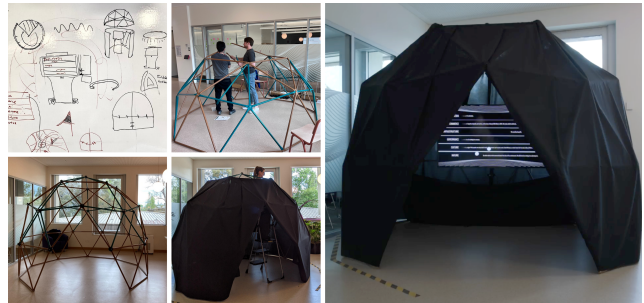

Figure 3. The images show the construction of the dome, from sketches and assembling to placing the custom-shaped black fabric to create a dedicated dimly lit space for interaction.

**Creating a Dedicated, Dimly Lit Dome for Exploration**
We constructed an organic-shaped geodesic dome from two dome climbers, creating a dedicated space for contemplation and gestural interaction. The dome measures 2.5 metres in height, 3.2 metres in diameter, and 1.5 metres in width at the entrance. The semi-translucent dark fabric was used to create a dimmed lit space, as this can elicit moments of recollection and introspection that "reintegrate the human psyche with the natural world" (Frey & Harper, 2023), and in an office context have shown to "increase freedom from constraints" (Steidle & Werth, 2013). The installation's ambience can alter office workers' perception of time and place. It invites engagement with diverse temporal scales, representing the variability and unpredictability of nature and human activity. This shift from computing experience for productivity aligns with creative computing focus on supporting wellbeing, curiosity, and connection to nature.

# User Study

**Participants Setting and Procedure**

We recruited 18 participants (7 Females, 8 Males, 3 Self-described), aged 22 to 56 years (Mean=39, SD=17.89). Who were recruited via advertising. A key requirement was that participants worked primarily in office settings. Participants did not receive any form of compensation. Ten participants were from our organisation—six from different departments and four from our department, none affiliated with the project or the research team. Eight participants were external. All worked at desks in various roles: 6 academics, 3 software engineers, 2 policy-law officers, 3 creatives, and 4 professional services staff. The installation is on the top level (level three) of a university building, in a corner of the foyer. It features a well-lit open space with large windows overlooking nature, allowing passers-by to hear the soundscape and watch interactions from behind.

Participants were individually interviewed at a table adjacent to the dome. In adherence to the best practices of the explicitation approach (Light, 1999; Vermersch, 1994), the interviewer positioned themselves not directly opposite but alongside the participant, both facing the dome. This setup was intended to support interviewees entering a state of evocation, where they often gaze into space while reliving and articulating their encounters with the technology. Before the interviews, participants were briefed on interacting with the installation through gestures. Our intention was to gain an understanding of their experience, impressions, and attitudes towards the installation to learn about how to design computing creative experiences for the office.

**Data Collection**

We use the explicitation interview approach (Bedin et al., 2019; Vermersch, 1994) to prompt participants' introspection across several "episodes" (Light, 1999) of technology encounters. Each of the three interview episodes comprised 8 minutes of self-guided technology exploration followed by 15 minutes of interviewing. We interviewed 18 volunteer office workers, resulting in 1530 minutes of interview audio recordings. Alongside observational notes and video recordings, this episodic approach helps participants move beyond the initial novelty of new technologies. It focuses on reflecting and articulating their latest experiences, allowing them to enter a state of evocation. Where they relive their encounters, providing rich details about their experiences.

To focus on specific episodes, we asked open-ended questions like: "As you entered the dome, what happened?", "What were you doing then?", "What was the installation doing?", "Where were you?", "Did this remind you of other experiences?", "What did you see and hear?", "How did you feel when this happened?", and "Tell me more about this." These prompted participants to relive their experiences.

**Data Analysis**

We employed an inductive thematic analysis approach to analyse the data (Clarke & Braun, 2014). The interview transcripts were coded independently by two authors using Nvivo software. This initial coding phase involved identifying prevalent themes in participants' descriptions of their experiences and conceptualisations of the use context.

Throughout this process, the authors held multiple discussions to refine their code books, using Miro to visually map out ideas and deepen understanding of emerging themes. Our analysis identified four comprehensive themes with nine subthemes from 298 coded units, supported by participant quotes and insights from designing the installation. The themes highlight participants' engagement and perceptions, contributing to our understanding of designing creative computing experiences for the office.

# Findings

The findings include four themes from 298 units coded.

**Theme one: Experiencing the office as a non-rationalised space (111 units)**

*A different spatial and visual reality. T1.1 (48 units)*

Participants described a different spatial reality when entering the dome. The dome's shape was characterised as *"a climbing playground"* or it *"reminds me of a planetarium"*. Participants felt the dome located them in places of play and wonderment, rather than their regular workplace. Other participants commented on the effect of the dark curtain around the dome, where the enclosure *"obscures the reality of the office"*, and *"limits your peripheral vision"*, such as in *"virtual reality [...] where you engage in a different reality"*. The sense of entering a non-workspace appears to open opportunities for curiosity and exploration.

*Altered perceptions of time. T1.2 (32 units)*

Participants noted that they lost a sense of bracketed, discrete time blocks typically experienced during the workday. Instead, they experienced *"timeless time, like being in a cinema"*. The sensation that *"you could be anywhere in the world at any time"*. The shift out of coordinated time into boundless time also meant that participants felt *"... you relax and open up your mind to a broader horizon"*, in direct contrast to work experiences that require focus on specific tasks. Such an experience of being outside the routine also came in terms of distancing from being inside the office itself, *"I zoomed out from my point of view to a higher level… you lose your sense of presence"*. These observations indicate that workers experienced an altered perception of time compared to the constant rhythm of office life.

*Experiences of rhythm, flow, and focus. T1.3 (31 units)*

Contrasting with typical office rhythms, the installation created unique experiences. One participant described it as *"steady pace fluctuation,"* while another likened it to visualising the *"building's living pulse"* in sync with nature's rhythms. This transitioned participants from routine office tempos to a rich sensory experience, deepening their connection to the audiovisual fluctuations. A participant described this as a *"totalising experience,"* highlighting their contemplative state. The sensation of immersion was viscerally noted, with a participant saying, *"I feel the data is

*passing through my body,"* another said, *"the darkness & immersive sound intensifies the sensations"* in the dome.

**Theme two (T2): A sense of immersion and teleportation beyond the office (79 units)**
*Entering a responsive environment. T2.1 (40 units)*
Participants reflected on the interaction as dynamic. One said, *"There's a constant sensing & acting within yourself and the installation."* Another noted the distinctiveness of the experience: *"I would tweak my choice of the word mirror. It was like engaging with another."* Insights relating to interaction were shared, *"I interact with the world on a large scale"* and *"You see where & how you sit in the relations,"* referring to the time scales, people, technology, and nature. These reflections differ from typical office small talk.

*Reorientations to experiences of nature T2.2 (39 units)*
Participants shifted out of habituating the day-to-day workplace and into the realm of the natural world. A participant said, *"The audiovisuals reorient you to the fact that you're in an ancient landscape"*, another noted, *"You locate the building on country [as referred by the local indigenous to the land, the intertwined connections and beliefs] in a way that is not often in your mind."* These comments revealed a holistic conceptualisation of the office, echoed by another participant, who said *"After playing, you become critical, what is it trying to tell me? And you see the building as part of larger web"*. Participants described an emerging world, *"I'm hearing insects and water sounds. It feels metaphoric all the different dynamics coming in"*, another said, *"It's like the god trick, I can see everything, but I'm nowhere"*, referring to a disembodied view to that of their office bound self, as described by Haraway (Haraway, 2013). The experience offered participants moments of connection to nature, away from the immediate experience of the office.

**Theme three (T3): Place-making from embodying the data (50 units)**
*Feeling an interconnected world-system. T3.1 (27 units)*
Participants shared insights on perceiving the office as a nexus of human, technological, and environmental connections. A participant said, *"You become part of the network... you're contributing to the way it behaves."* Others highlighted, *"The audiovisuals make you think about everything that goes into the building"* and *"We have blinkers on... forgetting what we're connected to and our actions' ripple effects."* Reflections on longer time scales beyond the office and human life span, suggest newfound awareness, *"The installation shows the intricate balance of forces. To plan sensibly for the future means recognising this complex system,"* and *"The installation is a tool for greater sensitivity to our world's complexity and our place within it."*

*New sensations and awareness of place. T3.2 (23 units)*
Participants reflected on the altered perception and understanding of the familiar office. A participant felt it *"strips away your notions of the office and invites you to think more deeply"*. Another noted, *"The sounds and text become one... that's the experience of this place."* The conversion of typically imperceptible data into audiovisuals reshaped participants' engagement with the space, fostering new attentiveness. A participant said, *"I have seen and heard the data... this brings a different awareness of this space."* Another noted, *"You pay attention to the sounds and time scales."*

**Theme four (T4): Ritualistic engagement and newfound agency. (58 units)**
*Zen-like moments in non-Zen space. T4.1 (34 units)*
Participants' interactions led to mindful moments. A participant said, *"The soundscape was like a pond, you have a water dropper, and you can only drop so much water at a time"*. Another noted, *"It's a delicate interaction as you pluck the strings"*. Many likened the gestural interaction to mindful practices, with one noting, *"It reminds me of Tai chi, you are moving with focus… the dots move with you"*, and another noting the feedback as it, *"gave me a flow to move."* These immersive, free-flowing, meditative experiences contrast with the typical linearity of office work.

*Exploring the boundaries of agency. T4.2 (24 units)*
The installation offered participants a sense of agency, a participant described this interaction as, *"it starts playfully... you explore moving the dots... then you focus on the sound and text relations"*. Another likened their influence to causing ripples in a large body of water, *"When you're moving around... your body causes the water to ripple."* This gave a perception of being part of a larger system, *"There is a world in which we... interact with larger scale, those are a small interaction with a very large thing"*. Some participants expressed a desire to influence the soundscape, *"I was trying to affect the soundscape... but I couldn't affect it"*. The installation, while allowing interaction, limited participant's extent of influence and ran in its own time, prompting one to say, *"It's pushed me to want to affect the data"*.

## Discussion and Implications for Creative Computing Experiences in Offices

Developing design knowledge for creative computing experiences that offer moments of temporal diversity and nature contemplation in the office presents challenges and opportunities. We discuss our user study findings and critically reflect as designer-researchers on some of the challenges and insights to support creative computing research and practice in exploring this design space.

**Ephemeral Qualities in Creative Computing Experiences Using Generative AI**
Our work with generative AI for creative computing experiences for office workers involves processing real-time data, including the time scale data streams and the weather API. This resulted in outputs that offered an ephemeral experience for office workers - referring to the transient audiovisual experience where each engagement episode was distinct (Döring et al., 2013b). This ephemeral quality is known to contribute to an expanded appreciation of time and reflects the constant flux of nature (Döring et al., 2013a).

Interestingly, participants reported experiencing a higher-level perspective that revealed relationships and temporal scales extending beyond their office life. This suggests that the installation made visible a broader web of relations through its audiovisual composition. This live composition is a dynamic property of the installation; in the material performance tradition (Barati et al., 2018; Schön, 1992), it refers to when the "material becomes active in disclosing the fullness of its capabilities, the boundaries between human and nonhuman performances are destabilised in productive practices". The installation's performance offers office workers a "reflective conversation" that unfolds through their interactions (Barati et al., 2018).

In working with generative AI, we embraced the inherent unpredictability and variability of the data streams. While we curated the physical installation, the interface representations, including text, motion, interaction and loosely defined audio attributes, we did not control the content itself. This allowed room for generative AI to act as performative material, resulting in an ephemeral audiovisual composition that invited curiosity and contemplation.

**Creative Computing Experiences as Practice Cultivation Opportunities**

The office worker exerts their agency on the installation, but through the installation's restricted tempo, audiovisual generation, and slow gestural interaction, the installation also exerts its agency onto the office worker. Previous works discussed material and non-human agency as a quality emerging through interaction, particularly in creative practices (Knappett & Malafouris, 2008). When the two agents come together a new hybrid agent is formed, allowing a human-nonhuman hybrid to become in the world (Shildrick, 2022).

In designing creative computing experiences that extend the perceptual abilities of office workers through interaction, there is a shift in the concept of interaction and experience within the office space. The deliberate pacing, sensory outputs, and restrictive interactive controls in the presented installation encourage a reflective and contemplative engagement, similar to creative practices that require focus and letting go of control. This mirrors how artists become attuned to the agentic properties of materials, for example clay, which changes over time and responds to the potter's movement, temperature, and environmental conditions (Malafouris, 2008). Consequently, creative computing experiences for office workers could serve as platforms for practice cultivation, known for their meditative benefits (Terzimehić et al., 2019; Williams, 2014), to enrich the daily experience of office workers.

**Creative Computing Experiences Can Amplify Ecological Temporalities**

Our installation departs from the dominant human-centric office worker technologies for productivity to one that acknowledges and fosters awareness about humans, technology, and nature world relations. The human-centric paradigm places humans at the top of a socio-cultural and ecological hierarchy (Andres, 2022; Arbib et al., 2022). This often sidelines the broader ecological impacts and the systemic interdependencies that define our relations.

Our installation contributes to more-than-human design (Tironi et al., 2023; Wakkary, 2021) and country-centred design (Abdilla, 2020), which envisions "more-than-human coexistence" as care for others including the environment. One of the challenges, in the context of office work, is going beyond conventional utility metrics. This requires technology designers and office workers to ask, and actively create space for, what type of creative computing experiences offices should offer and how we can design them to promote long-term sustainability and enhanced wellbeing benefits.

**Breaking Silos via Creative Computing Experiences**

With the growing ubiquity of technology in office environments, we are witnessing device saturation, as foreseen in the 90s by Weiser's work on ubiquitous computing (Weiser, 1993). While his ubiquity agenda called for calming and often invisible technologies for carefree work, Rogers progressed this agenda to call for technologies that instead offer "engaging user experiences" that "promote engaged living" (Rogers, 2006). Today's office space includes a range of technologies from personal wearables for health monitoring to laptops, tablets, smart whiteboards, and environmental controls regulating lighting and temperature. Although optimised towards an ideal office working state, these devices inundate us with notifications that fragment our attention. Additionally, they miss opportunities to promote office workers "engaged living", for example, by explicitly engaging with nature and connecting with varied time scales, which can offer mental relief from the rapid pace of work and encourage more contemplative perspectives.

As we continue to shape computing technologies in an office context, from AI optimisation narratives to human-AI productivity partnerships (Serbanescu et al., 2022), we must consider what we are optimising for and what is being lost as a result. How might we expand the narrative of technological optimisation to invite office workers to engage with their environments more meaningfully, incorporating elements of nature and temporal diversity? By incorporating designs that promote reflection on our natural surroundings and non-human temporalities, we can create office spaces that promote wellbeing, creativity, and a deeper connection to the broader world. Such an approach could lead to a more fulfilling work life, where technology enhances our understanding and appreciation of the world beyond the immediate demands of our work tasks.

## Conclusion and Future Work

We presented the design and user study of an installation for office workers, enabling moments of temporal diversity and connection to nature. The installation is a creative computing experience departing from the traditional focus on productivity. Our goal was to prototype and study creative computing in the office, delving into the human-technology relations to better understand what it might offer to office workers within the structured rhythms of office life.

Drawing on neuroscience insights and the slowing effects of nature sounds on time perception, we created an immersive, slow interaction, generative AI installation that composes an audiovisual space — serving as a perceptual portal into temporal realms beyond the linear rhythm of the office.

Our user study presents the lived experience of 18 office workers via explicitation interviews, observational notes and video recordings analysed through an inductive thematic analysis. Resulting in four themes, nine sub themes, and four design implications.

This work contributes to growing calls to design technologies that promote engagement with temporal diversity, enhancing ecological awareness (Hayles, 2012; Odom et al., 2022; Vallgårda et al., 2015), prioritise wellbeing and engaged living (Hook, 2018; Mueller et al., 2018; Rogers, 2006; schraefel et al., 2021), and contribute to narratives of human-machine hybrids to open possibilities for positive futures (Andres, Semertzidis, et al., 2023; Shildrick, 2022).

Future research could investigate the long-term effects on aspects like joy, interpersonal relationships, and connectedness. This would further our understanding of how to design, maintain, and decommission creative computing experiences. Future research could explore other senses, olfactory & tactile stimuli, building on human-flora interaction (Seow et al., 2022), to explore their benefits in offices.

Overall, our design and user study offer challenges and insights to support creative computing research and practice to enable enriching experiences in the office.

## Acknowledgments

Thank you to the participants for their time, the reviewers for their insightful feedback, and to Assoc/Prof. Oliver Bown for his contributions to this art piece.